\documentclass[article]{jfm}

\usepackage{graphicx}
\usepackage{newtxtext}
\usepackage{newtxmath}
\usepackage{natbib}
\usepackage{hyperref}
\hypersetup{
    colorlinks = true,
    urlcolor   = blue,
    citecolor  = black,
}

\newcommand{\RomanNumeralCaps}[1]

\title{A New Mechanism for Generation of Langmuir Circulations}

\author{Andre Basovich\aff{1}
  \corresp{\email{abasovich@cortana.com}},
  Dylan J. Wall\aff{2}
 \and Eric G. Paterson\aff{2}}

\affiliation{\aff{1}Cortana Corporation, Falls Church, VA 22222, USA
\aff{2}Virginia Polytechnic Insitute and State University, Blacksburg, VA 24061, USA}

\begin{document}
\maketitle

\begin{abstract}
A new mechanism has been identified that explains the generation of Langmuir circulations.  A wind-driven current in the presence of surface waves gives rise to an instability where the emerging circulations redistribute the turbulence in the cross-wind direction.  The non-uniform eddy-viscosity locally changes the rate of momentum transfer from the wind to the shear current, producing a non-uniform velocity field. The interaction of this non-uniform velocity field with the surface waves, due to the Craik-Leibovich vortex force, amplifies the circulations and creates a feedback mechanism.  The currently accepted CL2 model of instability assumes a constant eddy-viscosity.  This paper presents a model which explains the generation of Langmuir circulations and its predictions of both spatial and time scales are in good agreement with experimental results. The modeling approach combines a perturbation method with a RANS turbulence model. Through parametric variation of the perturbation, the growth rate and spatial scales of the circulations are extracted from the simulations.
\end{abstract}

\section{Introduction}
\label{sec:intro}
Langmuir circulations (LC), first described almost a century ago by \citet{langmuir1938}, is a well-known phenomenon that plays a very important role in the dynamics of the upper ocean layer. LC are periodic system of paired vortices oriented in the direction of the wind. They are generated as a result of an instability of a developing wind-driven drift current in the presence of surface waves. LC are observed at the surface as a system of streaks containing foam, floating particles, and organic films. The streaks are created by the converging transverse surface flow associated with counter-rotating circulations. The developing LC also change the current velocity in the direction of the wind and create a periodic system of areas with increased and reduced velocity, identified as jets and wakes. These jets and wakes coincide with convergence and divergence zones of the transverse flow at the surface created by LC and are located above the downwelling and upwelling zones in the water column. The change in the wind-driven current velocity is most significant at the surface and decreases with the depth. It has been observed on many occasions that emerging LC evolve from a smaller to larger spanwise (cross-wind) scales and, in general, the transverse scale of LC is close to the depth of a mixed layer \citep{leibovich1983,Smith1992}.  A correct LC model should describe these features of the phenomenon.  In this paper we consider LC that occur in natural bodies of water. There are experimental studies of the similar circulations in the laboratory experiments \citep{Melville1998}. These circulations have a different mechanism and, thus, are not discussed here. 

Existing theoretical models of LC are modifications of the Craik-Leibovich (CL2) model \citep{Craik1976,leibovich1983}. The CL2 model of LC introduced the vortex force: an average force imposed by surface waves on a wind-driven current with non-uniform velocity profile. This force was also introduced by \citet{garrett1976} and its expression was rigorously derived by \citet{Andrews1978}. Due to the effect of the vortex force, flow transverse to the current is generated and large circulations are produced by the interaction of the non-uniform current and surface waves.  Current models of instability describe the generation of circulations which lack many features of observed LC.  A detailed discussion of the discrepancy of the current theoretical predictions and results of the observations was given by \citet{Basovich2014}. The main problem with the CL2 model is that it does not produce the characteristic scale of the fastest growing LC, which is clearly observed in the ocean.  It is asserted that the CL2 model is unable to predict the characteristic scale due to the lack of a feedback mechanism that causes the instability. Further, the CL2 model predicts that the maximum change of the drift current velocity is below the surface rather than at the surface, in contrast to the experimental data. 

There are two models that can produce a characteristic
scale of the most unstable LC, but they too have deficiencies.
The first is a so-called generalized CL (CLg) model
proposed by \citet{phillips2005}.
It includes the potential effect of a non-uniform velocity
profile on the surface waves, and produces a characteristic
circulation scale.
However, as with the original CL2 model, CLg does not
correctly describe the modification of the shear profile 
by the circulations, and the estimates provided by the model
depend on an assumed distribution of the eddy-viscosity that is not justified. The second model \citep{Basovich2014} assumes that the presence of bubbles in the near-surface layer can affect the eddy viscosity in the layer. The eddy viscosity controls the boundary condition for velocity of the drift current at the surface. Although the second model can describe the main features of the LC, the estimates provided by the model also depend on an assumed value of eddy viscosity and cannot be easily compared to the experimental data. We believe that this mechanism can play an auxiliary role to the mechanism described here. 

\begin{figure}
    \includegraphics[width=0.9\textwidth]{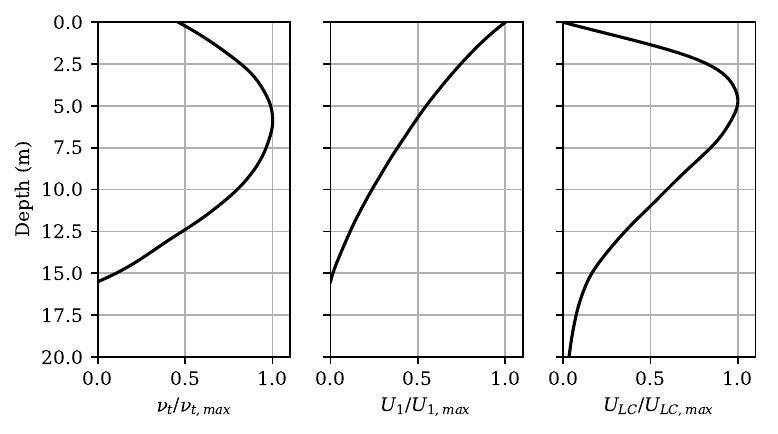}
    \caption{Normalized values of the horizontal
        velocity component of the shear profile
        ($U$), the eddy-viscosity ($\nu_t$),
        and the vertical velocity for the
        fastest-growing LC.
        Profiles were sampled at 20 minutes
        from onset of the wind with speed of
        $V_{10} = 13$ m/s.}
    \label{fig:prof}
\end{figure}

The statement of the problem of instability of the wind-driven drift current leading to the generation of LC presented in this paper is similar to the one used by the CL2 model. First, the development of the wind-driven drift current is described within a framework of the $k-\epsilon$ RANS  turbulence model. At some stage of development of the current, the vortex force imposed on the current by the surface waves is introduced and the instability of the current is analyzed. The principal difference between the new model (referred herein as BWP) and the CL2 model is that the CL2 model assumes the eddy viscosity of the wind-driven current to be constant, while the present $k-\epsilon$ model calculates the eddy-viscosity from turbulence model quantities. Figure \ref{fig:prof} shows the profiles of velocity and eddy-viscosity for a wind-driven shear layer developed using the $k-\epsilon$ model (see, for example, \citet{Burchard2001b}). The wind for the presented profiles had a speed at 10 m elevation of $V_{10}=13$ m/s, similar to the conditions measured by \citet{Smith1992}. The eddy-viscosity in such a current has its maximum below the surface. The value of eddy viscosity near the surface determines the transfer of momentum from wind to the current: when the near-surface eddy viscosity is lower, fluid accelerates faster by the wind and it accelerates slower when the eddy viscosity is higher. The LC arising in the drift current will increase the eddy viscosity of the fluid above the areas of the upwelling due to advection of the water with higher eddy viscosity upward. These areas correspond to divergence zones of the LC-related transverse flow at the surface and to the wakes in the drift current. The increase of eddy viscosity near the surface will decrease velocity of the drift current in these areas even further. The opposite process will take place above the areas of downwelling that correspond to convergence zones of transverse surface flow and jets in the drift current. This process will increase the non-uniformity of the drift current. The increase of the current non-uniformity increases the vortex force imposed on the current by surface waves (Stokes drift) and, thus, amplifies the initial LC. This is the feedback mechanism that drives the instability generating the LC.  

\section{Description of the model of instability}
\label{sec:model}
For this study, a series of numerical experiments
were performed. Each was conducted in two stages.
A wind-driven shear layer was simulated, then
a perturbation was introduced to the
velocity profile and the model vortex force was activated.
Although a realistic description of the development
of a mixed layer should be simulated simultaneously with
LC development, our focus here is on the physics of
the instability leading to generation of LC.
The simulations of the simultaneous
development of the current and LC will be discussed
later.

\subsection{Governing Equations}
\label{sec:goveqn}
The CL2 model assumes a constant eddy viscosity.
In contrast, the model employed for this study uses
the standard two-equation $k-\epsilon$ RANS turbulence
model as described by \citet{Launder1974}.
The development of the wind-driven mixed layer has been
previously described using the $k-\epsilon$ model
\citep{Burchard2001}.
A comparison of the predictions of different turbulence
models is presented by \citet{Burchard2001b}.
These studies demonstrated that the $k-\epsilon$ model,
modified for geophysical flows, can provide similar
predictions to more complicated turbulence model
closures such as the Mellor-Yamada model \citep{Mellor1982}.
The unsteady Reynolds-averaged Navier-Stokes
and continuity equations for incompressible
fluids are given below.    
\begin{align}\label{e:mntm}
    \frac{\partial U_i}{\partial t}
  + U_i  \frac{\partial U_i}{\partial x_j}
    =
  - \frac{1}{\rho} \frac{\partial p}{\partial x_i}
  + \frac{\partial}{\partial x_j} \left[
        \nu \frac{\partial U_i}{\partial x_j}
      + \nu_t \left(
            \frac{\partial U_i}{\partial x_j}
          + \frac{\partial U_j}{\partial x_i}
        \right)
      - \frac{2}{3} k \delta_ij
    \right]
  + \frac{1}{\rho} F^v_i,
\end{align}
\begin{align}
  \frac{\partial U_i}{\partial x_i} = 0,
\end{align}
where $U_i$ are components of the average velocity $\vec{U}$,
$p$ is pressure, $\rho$ is the water density,
$\nu$ is the kinematic molecular viscosity,
$\nu_t$ is the turbulent eddy-viscosity,
$k=\frac{1}{2} \overline{u_i u_j}$ is the turbulence
kinetic energy per unit mass,
and $u_i$ are fluctuations in the velocity components.
The coordinates $x_i$ are ($x$,$y$,$z$), with the $x$ axis
aligned with the wind, the $y$ axis normal to the wind,
and the vertical axis $z$  directed upward.
The last term in equation (\ref{e:mntm}) is the vortex
force imposed by surface waves on the
time-varying average velocity field: 
\begin{align}\label{e:Fv}
\vec{F}_v
=
\rho \vec{V}_s \times \vec{U},
\end{align}
where $\vec{V}_s=U_s \hat{x}_1$ is the Stokes drift
velocity.
For the stability analysis detailed here, the
long surface waves are represented as a simple
monochromatic wave in deep water, following
\citet{leibovich1983}.
The Stokes drift velocity is given by:
\begin{align}\label{e:Us}
U_s
=
(k_s a_s)^2 \sqrt{g/k_s} e^{2 k_s z},
\end{align}
where $k_s$ and $a_s$ are the wavenumber and amplitude
of the surface wave, and $g$ is the gravitational
acceleration.
\subsection{Turbulence Model}
The two-equation $k-\epsilon$ model of \citet{Launder1974}
is given below.
\begin{align}
    \frac{\partial k}{\partial t}
  + \frac{\partial(kU_i)}{\partial x_i}
    =
    \frac{\partial}{\partial x_i} \left[
        \nu
      + \frac{\nu_t}{\sigma_\kappa}
        \frac{\partial k}{\partial x_i} 
    \right]
  + \mathcal{P}
  - \epsilon,
\end{align}
\begin{align}
    \frac{\partial \epsilon}{\partial t}
  + \frac{\partial (\epsilon U_i)}{\partial x_i}
    =
    \frac{\partial}{\partial x_i} \left[
        \nu
      + \frac{\nu_t}{\sigma_\epsilon}
        \frac{\partial \epsilon}{\partial x_i}
    \right]
  + \frac{\epsilon}{k}
    \left( C_1 \mathcal{P} - C_2 \epsilon \right),
\end{align}
\begin{align}
\nu_t = C_\mu \frac{k^2}{\epsilon},
\end{align}
\begin{align}
\mathcal{P}
=
\nu_t \left(
    \frac{\partial U_i}{\partial x_k}
    +
    \frac{\partial U_k}{\partial x_i}
\right) \frac{\partial U_i}{\partial x_k}.
\end{align}
The model constants are
($C_1=1.44$, $C_2=1.92$, $C_\mu=0.09$) and the
transport constants are
($\sigma_k=1$, $\sigma_\epsilon=1.3$).
Note that in the initial 1D development of the wind-driven
mixed layer, the production term $\mathcal{P}$ simplifies
to:
\begin{align}
\mathcal{P}
=
\nu_t \left( \frac{\partial U}{\partial z} \right)^2.
\end{align}
The ocean surface boundary conditions for the velocity
are Dirichlet for the vertical 
component ($W = 0$) and Neumann for the 
wind-aligned and wind-normal components:
\begin{align}
    \frac{\partial U}{\partial z}
    =
    \frac{\tau}{\nu_t},
\end{align}
where $\tau$ is the shear stress applied by the wind on
the ocean surface.
The surface boundary conditions for the turbulence
quantities conform to a standard high Reynolds number
wall function:
\begin{align}
    \frac{\partial k}{\partial z} =\, 0,
\end{align}
\begin{align}
    \label{e:eps}
    \epsilon
    =
    \left(C_\mu\right)^3 \frac{k^{3/2}}{\kappa(z+z_0)},
\end{align}
where $\kappa=0.4$ is the K\'arm\'an constant, $z$ is
the wall-normal surface distance of the closest
numerical discretization point,
and $z_0$ is a surface roughness value.
\subsection{Initial Conditions and Perturbation}
In order to simulate the development of Langmuir cells
under conditions similar to the measurements of
\citet{Smith1992},
a two-stage simulation was employed.
First, the development of a wind-driven mixed layer and
the associated shear profile was simulated starting
from a quiescent state using a one-dimensional domain.
During the 1D simulation, the vortex force
term (\ref{e:Fv}) was inactive.
After the passage of a time $t_1$, the profiles for
$U_i$, $k$, and $\epsilon$ were interpolated to a
two-dimensional $(y,z)$ domain and used as initial conditions
for a second (2D) simulation.
The velocity profile was then modified by the application
of a small two-dimensional perturbation periodic in the
cross-wind direction:
\begin{align}\label{e:perturb}
    U (y,z)
    =
    \left[
    1
    +
    A \sin \left( \frac{2\pi y}{L} \right)
    \right]
    U_{t_1} (z),
\end{align}
where $U_{t_1} (z)$ is the unperturbed velocity
profile (such as depicted in Figure \ref{fig:prof})
and $A$ is a small coefficient $A\ll1$.

\section{Simulation of instability leading to generation of the LC}
\label{sec:simulation}
There are many potential environmental conditions which
might be considered in the study of LC generation.
An analytical study of the instability leading to the
appearance of LC under all conditions is not feasible
at this point.
Thus, a particular set of conditions was chosen that
reflect an experimental situation observed in the study
by \citet{Smith1992}.
A development of the drift current under the wind with
a speed of 13 m/s at 10 m elevation in the presence of
long surface waves is considered.
Initially, the vortex force is not included.
A surface roughness parameter in the boundary condition (\ref{e:eps})
is taken as 2 m, which corresponds to the amplitude of the
long surface waves that is on the order of 1 m.
The value of the wind speed corresponds to a wind stress of
$\tau=0.16 \, \mathrm{N/m^{2}}$.

Development of the wind-driven current from the onset of the wind is computed using the $k-\epsilon$ model described by a system of equations and boundary conditions (\ref{e:mntm}-\ref{e:eps}). Examples of profiles of the current velocity and eddy viscosity as a function of the depth are presented in Figure 1 for a current development time of 20 minutes from the onset of the wind. Such profiles of the wind-driven mixed shear layer are well known (see, for example, \citet{Burchard2001b}). The depth of the mixed shear layer in Figure 1 is approximately 15.5 m. The depth of the layer increases with time. The depth is 20 m for a development time of 30 min, and 24.5 m for a time of 40 min. The shapes of the velocity and eddy-viscosity profiles are similar to what is presented in Figure 1. The magnitude of the eddy-viscosity reaches maximum approximately in the middle of the layer due to a higher mixing length away from the surface. 

When a vortex force caused by the surface waves is imposed
on the wind-driven current, the current becomes unstable
and LC with vorticity aligned with the current arise.
The results of simulations of such a process using the BWP model are described here.
Note that if the development of LC is simulated from a completely
quiescent state, the structure of the growing LC is
irregular and depends on the characteristics of the
numerical discretization and numerical error.
Therefore, in order to make meaningful conclusions about
the growth rate of LC as a function of a spanwise scale,
LC with an arificially imposed scale $L$ were studied.
A very small perturbation to the velocity was imposed on the
drift current using the form of equation (\ref{e:perturb}).
This perturbation is not an exact eigenfunction of the
problem since this function is not known.
Therefore, it takes some time for the perturbation to form
a structure corresponding to eigenfunctions for all the
variables.
The perturbation then grows exponentially with time at a
constant growth rate $q$.
In the described simulations the time required for LC to
form such a structure is on the order of ten minutes.
As LC grow, the shear current also continues to develop.
Thus, we attribute the growth rate of LC to the time at
which it is being recorded.
For example, small perturbations are imposed on the current
ten minutes prior to the time when the growth rate of LC
becomes stable and is being recorded.
The computed values of the LC growth rate for each scale
$L$ fluctuate slightly from time step to time step.
Thus, these values are averaged over several time steps.
The behavior of the LC related change of the horizontal
velocity $U$ is similar to the profile of its initial
perturbation (\ref{e:perturb}).
The maximal value occurs at the surface, which is consistent
with the observations.   

Figure 2 shows the rate of growth of LC as a function of spanwise
scale $L$.
The curves are spline interpolations between the exponential growth
rates observed in multiple simulations with varying perturbation scales
$L$.
The results are presented for three stages of drift current
development, with $t_1 = 20$, 30, and 40 minutes.
The plot in Figure 2 demonstrates an existence of a characteristic
scale of the LC instability $L_0$ with the maximum rate of growth of
LC for each stage of the drift current development.
For the specified three time periods, the values of $L_0$ are
approximately 20 m, 27 m, and 30 m.
The characteristic times of the LC growth
(the inverse of the growth rate) are approximately
230 s, 290 s, and 305 s (between 3.5 and 5 minutes). 

\begin{figure}
    \centering
    \includegraphics{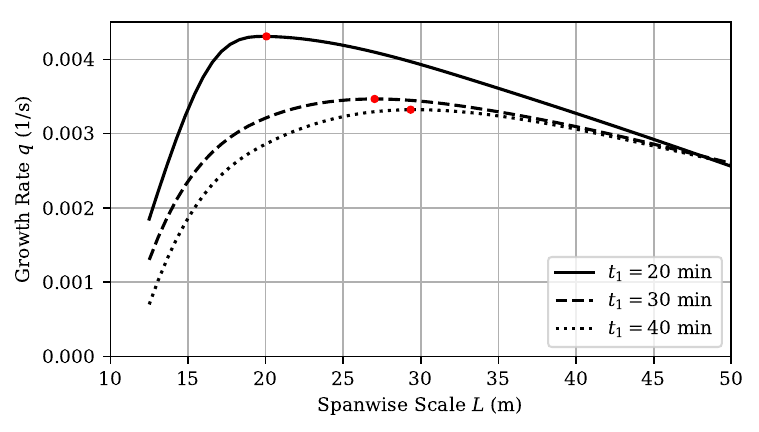}
    \caption{Circulation growth rate as a
        function of cross-wind perturbation
        scale for different times from wind
        onset, at wind speed $V_{10}=13$ m/s.
        Dots indicate the characteristic scale $L_0$.}
    \label{fig:growth_rate}
\end{figure}

The rate-of-increase of the scale $L_0$ with time as
the current develops is approximately 5 m per 10 minutes
or 30 m/h.
In all the cases simulated, the ratio of the characteristic
LC scale to the depth of the mixed layer is close to 1.3.
This ratio holds in simulations for different wind speeds.
The characteristic vertical scale of LC can be inferred
from the profile of the vertical component of the velocity
of the vortices for the fastest growing LC.
An example of such a profile of the velocity,
normalized by its maximum value, is shown in Figure 1
for the drift current development time of 20 minutes.
The area where the vertical velocity component of the
LC vortex is significant is roughly within the depth
of the drift current.
This relation is observed in simulations for all times
of development of the drift current at different wind
speeds.
Therefore, the characteristic vertical scale of the
generated LC is close to the depth of the shear layer.
For a wind speed of 13 m/s, the depths of the shear
layer for the time intervals of 20, 30 and 40 minutes
is approximately 15.5 m, 20 m, and 24.5 m.
Thus, the rate of growth of the depth is estimated as
4.5 meters per 10 minutes or 27 m/h.
The spanwise scale of LC corresponds to a pair of
vortices; thus, the aspect ratio of the fastest growing
LC vortex is half of the ratio of the characteristic
LC scale $L_0$ to its characteristic vertical scale.
This ratio is close to half of the ratio of the
characteristic LC scale to the depth of the mixed
layer and equals approximately 0.65.

The results described above are obtained for one particular set of environmental parameters. It should be noted that the characteristic scale and rate of growth of the LC depend on parameters of the surface waves. For example, an increase in amplitude of the surface waves will produce a higher rate of growth, while a gradual decrease in the amplitude will eventually suppress the instability.  Also, the effect of wave breaking was neglected, while in other situations this effect may be very important. 
A full parametric study of the development of LC under different environmental conditions is out of the scope of this short paper. Such a study will require significant improvement of the existing model and is discussed in the next section. However, the results obtained to date help explain the most important experimental results of observed LC growth.

\section{Comparison of results of simulations and observations}
\label{sec:comparison}
There are many reports of observations of LC, as reviewed by
\citet{leibovich1983}.
However, only the distinct paper by \citet{Smith1992}
contains information sufficient for the comparison of
simulation predictions and experimental results.
The observations described in that paper have gone largely
unexplained for three decades.
A potential explanation of these experimental results was
suggested by \citet{Basovich2014}.
Unfortunately, the predictions of the model presented in
that paper depend on many assumptions that are difficult to
verify.
A paper by \citet{Kukulka2009}, which reports
eddy-resolving simulations of Langmuir turbulence,
shows reasonable agreement with Smith's measurements
in the later stages of mixed layer development, when
the Langmuir cells contain significant chaotic motion
and a broad range of scales.
However, in the initial stages of development, (stage 2 as
presented by \citet{Kukulka2009}), the LES simulations do
not demonstrate the characteristic scale and
growth rate evident in the observations of \citet{Smith1992}.
In contrast, it is asserted that the model presented in this
paper describes the observed sequence of the events in this
initial stage with minimal assumptions.

First, recount the sequence of events
observed by \citet{Smith1992}.
It was reported by \citet{Smith1992} that a strong
wind was blowing for tens of hours and that long
surface waves were present.
At that time no LC were observed.
The wind then increased abruptly from 8 m/s to 13 m/s.
LC developed within approximately 20 minutes
of the onset of the wind increase.
Initially, relatively small-scale LC with a spanwise
scale of approximately 16 m appeared.
As time progressed, LC with increasingly large scales
were observed.

The rate of increase of the spanwise  scale $L_0$
(defined as a distance between the streaks) was estimated
by \citet{Smith1992} to be approximately 40 m/h.
The growth of LC with a well pronounced $L_0$ continued
for roughly 50 minutes.
After this growth interval, the circulation cells
devolved into a chaotic field of circulations with a
range of spanwise scales from 40 to 100 meters.
These LC continued for several more hours.

Subsequent to the generation of LC described above,
the wind speed gradually decreased to below 10 m/s and
about 2 hours later increased to 13 m/s again.
No new LC with a distinct characteristic scale were
observed after this second increase in wind speed.
At the onset of the first wind speed increase and
appearance of LC, the top 15 meters of the upper layer
of the ocean were weakly stratified.
Monitoring of the water density showed intensive mixing
of the upper layer that only can be attributed to LC.
The depth of the mixed layer increased at a rate of 20 m/h.
This depth is also used as an estimate of a vertical scale
of the developing LC.  From the comparison of the rates of
increase of the spanwise scale (that corresponds to a pair
of vortices) and vertical scale the aspect ratio of the LC
vortex was estimated to be close to one \citep{Smith1992}.

Three observations that require an explanation are listed below.
First, at the beginning of the generation process, the LC have a
well-pronounced spanwise scale that increases with time.
Second, LC are not observed before the increase in the wind speed,
even though a strong wind had been blowing for tens of hours.
Third, the generation of new LC was not observed after the second
increase of the wind.   

The simulation predictions presented herein describe
qualitatively all the major features of the initial
stages of of LC generation observed by \citet{Smith1992},
and suggest explanations for the observations listed above.
A quantitative comparison of the simulation results with
experimental data is also conducted for the initial stage
of LC development.
At the initial stage of the LC growth
(with a wind speed of 13 m/s) the simulations demonstrate
the appearance of LC with spatial and time scales very close
to what was reported by \citet{Smith1992}.

The rate of increase of the spanwise Langmuir cell scale in
the simulations associated with the characteristic scale
$L_0$ was approximately 30 m/h.
The rate estimated by \citet{Smith1992} for the horizontal
scale was 40 m/h.

The explanation for the discrepancy in this case is
subtle, and requires an understanding of the last stage
of the evolution of the observed circulations.
It was observed by \citet{Smith1992} that after
approximately 50 minutes of
development there is not a system of cells with a single
pronounced spanwise scale, but rather a wide spectrum of scales is present.
At the same time the deepening of the mixed layer stops.
At this stage strong LC are developed and the upper layer
is fully mixed.
The rate of increase of the characteristic scale of LC is
estimated by \citet{Smith1992} based on the whole interval
of observation, including the later stages with significantly
more chaotic motion.
However, the estimate of the growth rate from the
presented simulation predictions was obtained only during
the initial growth of LC with a well pronounced scale $L_0$.
In this interval of 50 minutes, which takes place from time
marked as 7.6 hours to 8.5 hours
(see Figure 7 in \citet{Smith1992}),
the scale changes by approximately 25 m (from 15 m to 40 m).
Estimating the growth rate using just this initial interval
from Smith's data gives a growth rate of $L_0$ of 30 m/h,
which is in agreement with the presented simulation results.

Similarly, in the simulation associated with the characteristic
scale $L_0$, the rate of increase of the vertical Langmuir cell scale
in our simulations was 27 m/h. 
The rate estimated by \citet{Smith1992} under similar conditions
was 20 m/h. In this case, the discrepancy is readily explained
by the density stratification, which was omitted from the
simulations detailed in this paper.

Approximately at the time marked as 8.5 hours (as seen in Plates 3 and 4 in the paper by \citet{Smith1992}) the deepening of the mixed layer stops and the LC no longer show a dominant scale. This occurs because, the velocity and eddy viscosity profiles presented in Figure 1 are altered due to the intense mixing caused by the developing LC.  When the upper layer is well mixed, these values no longer depend on depth. Correspondingly, there is no instability related to the change of the eddy-viscosity, and the generation of the LC stop. However, the already developed LC can exist for a long time due to the large scale of the vortices; therefore, the mixing continues.  At that time a wide spectrum of spanwise scales of the perturbations of a near-surface current are observed with scales of up to 100 m. The presence of these large scales in the surface current does not necessarily mean that there are LC vortices with such scales. Our simulations show that the large-scale LC would have a comparable vertical scale. Since there is no deep mixing observed at this stage, we conclude that there is a chaotic system of the residual LC-like vortices with a characteristic scale on the order of the depth of the mixed layer (about 30 m). This system continues to evolve, and these vortices continue to get energy from the wind and waves as long as some non-uniform current at the surface exists \citep{Basovich2011}. This surface current structure has a wide spectrum with different spanwise scales that are observed in the experiment \citep{Smith1992}.

Two additional features of the observations of
\citet{Smith1992} can also be explained by
appealing to the instability mechanism presented here.
First, in Smith's experiment no newly generated LC
were observed at the second strong increase of
wind speed because residual LC continued the
process of mixing.
Advective mixing prevented formation of the shear
layer with a structure described in Figure 1,
which is the condition for the instability and
generation of new LC.
Second, the reason for the absence of LC before
the significant increase in the wind speed after
a long interval of blowing wind is similar:
the upper layer was well mixed during that
time interval and the eddy viscosity was nearly
uniform in the upper layer.
This qualitative description of the late stage
of evolution of LC explains the processes
that take place.

One of the advantages of the present model is that its results
can be easily reproduced, since it consists of a standard
$k-\epsilon$ model with the addition of the vortex force
and a modified sea-surface boundary condition for the
velocity gradient of the drift current.
However, the $k-\epsilon$ model has well-known limitations
when applied to problems with recirculating flow or
significant streamline curvature. 
The predictions of the model reported here should therefore
be trusted only for the initial stage of the instability
leading to the generation of LC.
The growth of the transverse velocity produced by LC
makes the structure of the layer no longer planar and
reduces the reliability of $k-\epsilon$ model predictions.

Several modifications might be made to improve the model
predictions in light of the discrepancies noted above;
the most significant are the introduction of
stratification effects, and better
accommodation of streamline curvature.
A RANS description of the later stages of growth,
which include more chaotic motions often termed Langmuir
turbulence, will likely require additional model production
terms. 

Ultimately, the inception and growth of LC is
a very unsteady process, and eddy-resolving
approaches will provide more insight than
the RANS approach detailed here.
Quantitative study of these important processes
might therefore be improved with
LES (Large Eddy Simulations).
LES have been used in study of the chaotic system
of LC called Langmuir turbulence
\citep{mcwilliams1997,Belcher2012}.
Unfortunately, these simulations have two flaws which might reduce the reliability of their
predictions.
First, the key to the instability mechanism proposed
in this paper is the modification of the rate of momentum
transfer from the wind and surface waves due to the
redistribution of near-surface eddy-viscosity.
It is unclear that the combination of resolved scale
eddies and sub-grid scale viscosity models used by
the typical LES can correctly reproduce this effect,
particularly under the assumption of a rigid lid.
Second, it is asserted that the way the vortex force
is applied to the flow in most LES studies is
incorrect.
This force is averaged over many periods of the surface
waves and can be applied only to the movements of the
fluid with characteristic times much larger than a
period of longest waves.
However, in the equations used in LES of
Langmuir turbulence, this force is applied to the
movements of the fluid with all the time scales.
Therefore, the LES-based description of a system of
chaotic LC and its role in the formation of the upper
mixed layer of the ocean should be revisited.

\backsection[Acknowledgements]{Mr. K.J. Moore, President of Cortana Corporation, and Dr. John Pierce are acknowledged for their continuous support and fruitful discussions.   Dr. Jonathan Pitt is recognized for his technical contributions to our research group and his editorial review of the manuscript.}

\bibliographystyle{jfm}
\bibliography{jfm}

\end{document}